\documentclass[a4paper,fleqn,usenatbib]{mnras}
\usepackage[T1]{fontenc}
\usepackage{ae,aecompl}
\usepackage{graphicx}
\usepackage{amsmath}
\usepackage{amssymb}
\usepackage{txfonts}

\def\robert#1{{ #1}}

\newcommand{\ltsima}{$\; \buildrel < \over \sim \;$}
\newcommand{\lsim}{\lower.5ex\hbox{\ltsima}}
\newcommand{\gtsima}{$\; \buildrel > \over \sim \;$}
\newcommand{\gsim}{\lower.5ex\hbox{\gtsima}}
\newcommand{\bra}{\left\langle}
\newcommand{\ket}{\right\rangle}

\newcommand{\dd}{\mathrm{d}}

\newcommand{\ci}{\mathrm{i}}

\newcommand{\chip}{{\chi^\prime}}
\newcommand{\chipp}{{\chi^{\prime\prime}}}

\newcommand{\veck}{\bmath{k}}

\title[Local Universe influence on surveys]
{Influence of the local Universe on weak gravitational lensing surveys}
\author[R. Reischke, B.M. Sch{\"a}fer, K. Bolejko, G.F. Lewis, M. Lautsch]
{Robert Reischke$^{1,2,3}$\thanks{e-mail: r.reischke@technion.ac.il}, Bj{\"o}rn Malte Sch{\"a}fer$^1$\thanks{e-mail: bjoern.malte.schaefer@uni-heidelberg.de}, Krzysztof Bolejko$^{4,5}$,\newauthor Geraint F. Lewis$^5$,  Max Lautsch$^{1}$\\
$^1$Astronomisches Rechen-Institut, Zentrum f{\"u}r Astronomie der Universit{\"a}t Heidelberg, Philosophenweg 12, 69120 Heidelberg, Germany\\
$^2$Institut f\"ur Kernphysik, Karlsruher Institut f\"ur Technologie, 76344 Eggenstein-Leopoldshafen, Germany\\
$^3$Physics Department, Technion, 3200003 Haifa, Israel\\
$^4$School of Natural Sciences, College of Sciences and Engineering, University of Tasmania, Private Bag 37, Hobart TAS 7001, Australia\\
$^5$Sydney Institute for Astronomy, School of Physics, A28, The University of Sydney, NSW, 2006, Australia}

\begin{document}
\onecolumn

\pagerange{\pageref{firstpage}--\pageref{lastpage}}
\pubyear{2018}
\maketitle
\label{firstpage}

\begin{abstract}
Observations of the large-scale structure (LSS) implicitly assume an ideal FLRW observer with the ambient structure having no influence on the observer. However, due to correlations in the LSS, cosmological observables are dependent on the position of an observer. We investigate this influence in full generality for a weakly non-Gaussian random field, for which we derive expressions for angular spectra of large-scale structure observables conditional on a property of the large-scale structure that is typical for the observer's location. As an application, we then apply to the formalism to angular spectra of the weak gravitational lensing effect and provide numerical estimates for the resulting change on the spectra using linear structure formation. For angular weak lensing spectra we find the effect to be of order of a few percent, for instance we estimate for an overdensity of $\delta =0.5$ and multipoles up to $\ell=100$ the change in the weak lensing spectra to be approximately 4 percent. We show that without accounting for correlation between the density at observer's location and {\robert{the}} weak gravitational lensing spectra, the values of the parameters $\Omega_m$ and $\sigma_8$ are underestimated by a few percent. Thus, this effect will be important when analysing data from future surveys such as Euclid, which aim at the percent-level precision. The effect is difficult to capture in simulations, as estimates of the number of numerical simulations necessary to quantify the effect are high.
\end{abstract}

\begin{keywords}
gravitational lensing: weak -- dark energy -- large-scale structure of Universe.
\end{keywords}

\section{Introduction}
Weak gravitational lensing is a mapping of the large-scale distribution of matter: Light bundles from distant galaxies appear sheared due to the differential deflection caused by gravitational potentials that the light bundle encounters on its way from the source to the observer \citep{kaiser_weak_1992, 2001PhR...340..291B}. This cosmic shear signal is an excellent probe of cosmological parameters, of the particle content of the cosmological model, and of gravity through its influence on the expansion dynamics of the Universe and the formation of cosmic structures \citep{huterer_weak_2002, huterer_weak_2010, amendola_measuring_2008}. The weak gravitational lensing signal can be quantified in terms of shear spectra \citep{hu_power_1999, jain_cross-correlation_2003} for Gaussian and bispectra \citep{cooray_weak_2001} for non-Gaussian statistics, and its sensitivity can be boosted using tomographic methods or three-dimensional decompositions, with control over systematics \citep{heymans_potential_2006, kitching_scale-dependent_2016}.

In the interpretation of large-scale structure data such as weak lensing there seems always to be an implicit assumption of idealised FLRW-observers, who are allowed to choose spherical coordinates centered on their position, with the consequence that the measured redshifts correspond to the cosmological ones with small corrections, and that the structures, in which the observers reside, have no influence on the observation apart from causing these small corrections. But due to correlations in the cosmic density field, structures are not statistically independent, implying that real observers, who are necessarily linked to overdense regions because galaxies form there, are likely to see biased correlations. This idea motivated us to construct conditional correlation functions, i.e. correlations of a cosmological observable that depends on the density field value a the observer's location. In this sense, our notion is that of an ensemble of observers, whose density corresponds to the density of galaxies and ultimately to the density of matter, if one assumes a straightforward linear biasing model for simplicity. Observations averaged over this ensemble will not be equal to the volume average (i.e. by placing observers at random positions in the large-scale structure and averaging over the considered volume). In particular with reference to the weak lensing effect, one would expect in addition that galaxies in the local Universe are not randomly oriented but are themselves intrinsically aligned in the local superstructure \citep{flin_orientation_1990}.

In this paper, we aim to provide an estimate by how much cosmic shear spectra obtained in gravitational lensing would be influenced by value of the density field at the observer's location in the large-scale structure. For that purpose, we assume that the observer is situated in a Gaussian or weakly non-Gaussian random field and derive conditional correlation functions, which marginalised over all possible density field values, will revert to the conventional volume average. Although we formulate the formalism with the density at the observer's location as a condition, any property that shows a nonzero correlation with the fundamental field that causes the signal can be such a condition.

We emphasise that the effect that we are investigating is purely related to correlations between the structures that are observed and the structure in which an observer is residing. First order effects would include the peculiar motion of the observer, which impacts on the measured redshifts at the level $\upsilon/c\simeq10^{-3}$ as measured by the CMB-monopole \citep{mather_early_1991}. Other physical influences of the observer's structure or state of motion are in reality present as well, and are known to cause effects on the observed large-scale structure signal \citep{amendola_peculiar_2008,bonvin_effect_2008, amendola_measuring_2011}. Firstly, the overdensity at the observer's location would cause a Sachs-Wolfe effect decreasing the photon's redshifts and would lead to an underestimation of the distances, or might introduce an integrated Sachs-Wolfe distortion depending on the local structure formation dynamics \citep{cooray_did_2005, rakic_microwave_2006, maturi_actual_2007, francis_estimate_2010, notari_measuring_2012}. In parallel, there can be gravitationally induced effects on magnification \citep{duniya_large-scale_2016}. Typically, this is of the order $\Phi/c^2\simeq (\upsilon/c)^2\simeq 10^{-7}$, and in magnitude comparable to the local Sunyaev-Zel'dovich effect that is present in the cosmic microwave background \citep{dolag_imprints_2005}. Secondly, effects associated to statistical isotropy breaking of the large-scale structure signal due to peculiar motion of the observer through relativistic aberration and a Doppler-contribution to the measured redshifts \citep{sereno_aberration_2008, chluba_aberrating_2011, kosowsky_signature_2011,  mertens_effect_2013,2018CUESTA}. They scale with $\upsilon/c\simeq10^{-4}$ and enter the spectrum quadratically, which causes changes to the spectra at a level typically of $10^{-8}$, although for low-redshifts and wide angles it may be slightly higher \citep{2014MNRAS.443.1900B}. Thirdly, large-scale tidal fields around the observer can have the effects of both changing the boundaries of the survey in terms of redshift in an anisotropic way and fourthly, are associated with large-scale modes of the matter distribution that modulate the mean density of matter inside the survey, giving rise to supersample covariance \citep{akitsu_super-survey_2016, li_galaxy_2017}. The last two effects become small if one approaches the horizon-scale, because there the FLRW-symmetries are effective and deviations from that necessarily small. These effects show that it is very fortunate that the Universe is structured on scales smaller than the horizon and that the perturbation are small, such that the FLRW-dynamics is recovered on large scales and the assumption of a comoving FLRW-observer with small corrections is good.
{\robert{Additionally, on larger scales higher order effects arising from relativistic contributions become important which include for example couplings between the gravitational potential at source and lens, time delay effects coupling to the lens. Further terms arise due to non-linear couplings to tensor and vector modes. These effects have been described extensively in \citet{Bernardeau2010} and can cause biases on there own if not included \citep[e.g.][]{2018Lorenz}. However, in this work we assume that these effects can be modelled correctly and thus will not influence the power spectrum.}}

The structure of the paper is following: Sec.~\ref{sect_cosmology} presents a summary of cosmology and weak gravitational lensing; Secs. \ref{sect_statistics} and~\ref{sect_implementation} outline the statistical method, including the conditional correlation function, and investigate the extend to which observable weak lensing spectra and bispectra depend on the location of the observer; Sec. \ref{sect_summary} summarises and discusses the results. Throughout the paper we assume the fiducial cosmological model to be a spatially flat $w$CDM-model, with specific parameter choices $\Omega_m = 0.3$, $n_s = 0.963$, $\sigma_8 = 0.834$, $h=0.678$. In addition we assume homogeneous and isotropic statistical properties of the cosmological large-scale structure and assume that the distribution of perturbation is close to the Gaussian statistic and can be captured by the perturbation theory in the lowest nonlinear order.

\section{cosmology, cosmic structures and weak gravitational lensing}\label{sect_cosmology}
Gaussian fluctuations of a statistically homogeneous and isotropic random field $\delta$ is described by the CDM-spectrum $P(k)$,
\begin{equation}\label{eq:powerspectrum}
\bra \delta(\bmath{k})\delta(\bmath{k}^\prime)\ket = (2\pi)^3\delta_D(\bmath{k}+\bmath{k}^\prime)P(k),
\end{equation}
which is normalised to the variance $\sigma_8^2$ smoothed to the scale of $8~\mathrm{Mpc}/h$,
\begin{equation}
\sigma_8^2 = \int_0^\infty\frac{k^2\dd k}{2\pi^2}\: W^2(k8\mathrm{Mpc}/h)\:P(k),
\end{equation}
with a Fourier-transformed spherical top-hat $W(x) = 3j_1(x)/x$ as the filter function.

Small fluctuations, $\delta^{(1)}$, in the distribution of dark matter grow, as long as they are in the linear regime $-1\ll\delta^{(1)}\ll +1$, according to the growth function $D_+(a)$ \citep{2003MNRAS.346..573L,1998ApJ...508..483W},
\begin{equation}
\frac{\dd^2}{\dd a^2}D_+(a) +
\frac{2-q(a)}{a}\frac{\dd}{\dd a}D_+(a) -
\frac{3}{2a^2}\Omega_m(a) D_+(a) = 0,
\label{eqn_growth}
\end{equation}
where the deceleration parameter is given in terms of the logarithmic derivative of the Hubble function, $2-q(a) = 3+\dd\ln H/\dd\ln a$. In addition, adiabaticity implies $\Omega_\mathrm{m}(a)/\Omega_{m0} = H_0^2/a^3/H^2(a)$. Within this approximation of linear structure growth, Gaussianity of the density field is conserved and inherited to observables that depend on the density field in a linear way.

Nonlinear structure formation in the regime $|\delta|\simeq 1$ causes the density field to acquire non-Gaussian statistical properties. In the lowest order of perturbation theory there is a quadratic contribution $\delta^{(2)}(\veck)\propto D_+(a)$ to the linear solution $\delta^{(1)}(\vec k)$, which can be shown to be \citep{Bernardeau2002} 
\begin{equation}
\delta^{(2)}(\veck) =
\int\frac{\dd^3k^\prime}{(2\pi)^3}\:
F_2(\veck^\prime,\veck-\veck^\prime)\:\delta^{(1)}(\veck^\prime)\delta^{(1)}(\veck-\veck^\prime),
\end{equation}
here the kernel is given by
\begin{equation}\label{eq:mode_coupling_f2}
F_2(\veck,\veck^\prime) =
\frac{10}{7} + \left(\frac{k}{k^\prime} + \frac{k^\prime}{k}\right)\mu + \frac{4}{7}\mu^2,
\end{equation}
with $k = |\veck|$, $k^\prime = |\veck^\prime|$ and $\mu = \veck\veck^\prime/(k k^\prime)$. In this limit, the bispectrum $B(\bmath{k},\bmath{k}^\prime,\bmath{k}^{\prime\prime})$ of the density field is given by
\begin{equation}
\bra\delta(\bmath{k})\delta(\bmath{k}^\prime)\delta(\bmath{k}^{\prime\prime})\ket =
(2\pi)^3\delta_D(\bmath{k}+\bmath{k}^\prime+\bmath{k}^{\prime\prime})B(\bmath{k},\bmath{k}^\prime,\bmath{k}^{\prime\prime}),
\end{equation}
where we abbreviated:
\begin{equation}
B(\bmath{k},\bmath{k}^\prime,\bmath{k}^{\prime\prime}) = 
F_2(\bmath{k},\bmath{k}^\prime)P(k)P(k^\prime) +
F_2(\bmath{k}^\prime,\bmath{k}^{\prime\prime})P(k^\prime)P(k^{\prime\prime}) + 
F_2(\bmath{k}^{\prime\prime},\bmath{k})P(k^{\prime\prime})P(k),
\end{equation}
such that the CDM-spectrum $P(k)$ grows $\propto D_+^2(a)$ and the bispectrum $\propto D_+^4$.

Because the gravitational lensing convergence $\kappa$ provides a linear mapping of the density field $\delta$, $\kappa$ inherits the non-Gaussian properties from $\delta$ and becomes itself non-Gaussian. It should, however, be noted that the non-Gaussianities of the density field get partially wiped off, due to the broad lensing kernel and the summation of {\robert{different modes of}} $\delta$. As the effects that we are considering are strongest on large scales, a perturbative description of nonlinear structure formation to second order should be sufficient: We will therefore restrict ourselves to angular scales $\ell\leq100$.

Weak gravitational lensing by the cosmic large-scale structure refers to the distortion of light bundles that reach us from distant galaxies \citep[for reviews, see][]{2001PhR...340..291B, bartelmann_topical_2010, hoekstra_weak_2008, kilbinger_cosmology_2015}. As the effect can be computed without any assumptions apart from gravity, it is an excellent probe of the evolution large-scale structure with its dependence on the cosmological model \citep{takada_power_2013-1,refregier_weak_2003,munshi_cosmology_2008}. 
The weak lensing convergence $\kappa$ provides a weighted line-of-sight average of the matter density $\delta$ if the gravitational theory is Newtonian in the weak-field limit,
\begin{equation}
\label{eq:lens_mapping}
\kappa = \int_0^{\chi_H}\dd\chi\: W(\chi)\delta,
\end{equation}
with the weak lensing efficiency $W(\chi)$ as the weighting function,
\begin{equation}
W(\chi) = \frac{3\Omega_m}{2\chi_H^2}\frac{1}{a}G(\chi)\chi,
\mathrm{~with~}
G(\chi) =
\int_\chi^{\chi_H}\dd\chi^\prime\:n(z)\frac{\dd z}{\dd\chi^\prime}\frac{\chi^\prime-\chi}{\chi^\prime}.
\end{equation}
$n(z)$ denotes the redshift distribution of the lensed background galaxies,
\begin{equation}\label{eq:srd}
n(z) = n_0\left(\frac{z}{z_0}\right)^2\exp\left(-\left(\frac{z}{z_0}\right)^\beta\right)
\quad\mathrm{with}\quad \frac{1}{n_0}=\frac{z_0}{\beta}\Gamma\left(\frac{3}{\beta}\right),
\end{equation}
and is the expected distribution of lensed galaxies of the Euclid-mission with a parameter $\beta=3/2$. $z_0$ has been chosen such that the median of the redshift distribution is 0.9. The Hubble function $H(a)=\dot{a}/a$ for a Friedmann-Lema{\^i}tre-cosmology with matter density $\Omega_m$ and dark energy density $1-\Omega_m$ is given by
\begin{equation}
\frac{H^2(a)}{H_0^2} = \frac{\Omega_m}{a^{3}} + \frac{1-\Omega_m}{a^{3(1+w)}},
\end{equation}
and defines the comoving distance $\chi$ as a function of scale factor $a$ through
\begin{equation}
\chi = -c\int_1^a\:\frac{\dd a}{a^2 H(a)},
\end{equation}
where the Hubble distance $\chi_H=c/H_0$ sets the distance scale for cosmological distance measures. Although current weak lensing surveys are tomographic, we present our results without any subdivision of the galaxy sample in redshift. However, they  generalise to tomographic and even to the 3-dimensional case. Furthermore, we will use convergence statistics instead of shear, {\robert{even though the latter is the main observable in weak lensing (independent techniques include magnification which can be measured as well using the distribution of galaxy magnitudes and sizes \cite{schmidt2012_detection})}}. Apart from the practical reason that it is much easier to deal with the scalar convergence $\kappa$ instead of the tensorial shear $\gamma = \gamma_+ + \ci\gamma_\times$, the statistics of the two are identical under the assumption of weak gravity and with the Born-approximation in place.

\section{Conditional random fields and their spectra}\label{sect_statistics}

\subsection{Conditional correlation functions}
We start by establishing a probability distribution $p(\kappa(\bmath{\theta}),\kappa(\bmath{\theta}^\prime),\delta)$ for the weak lensing signals $\kappa(\bmath{\theta})$ and $\kappa(\bmath{\theta}^\prime)$ along two lines of sight $\bmath{\theta}$ and $\bmath{\theta}^\prime$ and incorporate into this distribution a property that characterises the observer's location, i.e. we are seeking for the joint probability of the three random variables. Note that the random variables are of course correlated, since they all originate from the evolved density contrast random field, whose statistic is known perturbatively, especially on large scales this approximation is still well suited, since no shell-crossing occurs on these scales. 
For simplicity, we consider this property to be the value of the density field $\delta$, but this can be in principle any property of the cosmic large-scale structure which is typical for a specific observer, for example the value of the peculiar velocity field. From this probability density one can derive the expectation value of the two-point correlation $\bra\kappa(\bmath{\theta})\kappa(\bmath{\theta}^\prime)|\delta\ket$  conditional on a chosen value of $\delta$,
\begin{equation}\label{eq:spectra}
\bra\kappa(\bmath{\theta})\kappa(\bmath{\theta}^\prime)|\delta\ket =
\int\dd\kappa(\bmath{\theta})\int\dd\kappa(\bmath{\theta}^\prime)\:\kappa(\bmath{\theta})\kappa(\bmath{\theta}^\prime)\:p(\kappa(\bmath{\theta}),\kappa(\bmath{\theta}^\prime),\delta).
\end{equation}
Analogously, the same procedure can be applied to yield a three-point correlation function $\bra\kappa(\bmath{\theta})\kappa(\bmath{\theta}^\prime)\kappa(\bmath{\theta}^{\prime\prime})|\delta\ket$ conditional on the observer's position,
\begin{equation}\label{{eq:bispectra}}
\bra\kappa(\bmath{\theta})\kappa(\bmath{\theta}^\prime)\kappa(\bmath{\theta}^{\prime\prime})|\delta\ket =
\int\dd\kappa(\bmath{\theta})\int\dd\kappa(\bmath{\theta}^\prime)\int\dd\kappa(\bmath{\theta}^{\prime\prime})\:\kappa(\bmath{\theta})\kappa(\bmath{\theta}^\prime)\kappa(\bmath{\theta}^{\prime\prime})\:p(\kappa(\bmath{\theta}),\kappa(\bmath{\theta}^\prime)\kappa(\bmath{\theta}^{\prime\prime}),\delta),
\end{equation}
from the probability distribution $p(\kappa(\bmath{\theta}),\kappa(\bmath{\theta}^\prime)\kappa(\bmath{\theta}^{\prime\prime}),\delta)$. In either case, the idealised correlation functions $\bra\kappa(\bmath{\theta})\kappa(\bmath{\theta}^\prime)\ket$ and $\bra\kappa(\bmath{\theta})\kappa(\bmath{\theta}^\prime)\kappa(\bmath{\theta}^{\prime\prime})\ket$ would be obtained by marginalisation over the values of $\delta$ under the assumption of vanishing correlations between $\kappa(\bmath\theta)$ and $\delta$,:
\begin{equation}
\bra\kappa(\bmath{\theta})\kappa(\bmath{\theta}^\prime)\ket =
\int\dd\delta\: \bra\kappa(\bmath{\theta})\kappa(\bmath{\theta}^\prime)|\delta\ket =
\int\dd\delta\:p(\delta)\int\dd\kappa(\bmath{\theta})\int\dd\kappa(\bmath{\theta}^\prime)\:\kappa(\bmath{\theta})\kappa(\bmath{\theta}^\prime)\:p(\kappa(\bmath{\theta}),\kappa(\bmath{\theta}^\prime)) ,
\end{equation}
and, using again 
\begin{equation}
\bra\kappa(\bmath{\theta})\kappa(\bmath{\theta}^\prime)\kappa(\bmath{\theta}^{\prime\prime})\ket =
\int\dd\delta\: \bra\kappa(\bmath{\theta})\kappa(\bmath{\theta}^\prime)\kappa(\bmath{\theta}^{\prime\prime})|\delta\ket 
= \int\dd\delta\:p(\delta)\int\dd\kappa(\bmath{\theta})\int\dd\kappa(\bmath{\theta}^\prime)\int\dd\kappa(\bmath{\theta}^{\prime\prime})\: \kappa(\bmath{\theta})\kappa(\bmath{\theta}^\prime)\kappa(\bmath{\theta}^{\prime\prime})\:p(\kappa(\bmath{\theta}),\kappa(\bmath{\theta}^\prime),\kappa(\bmath{\theta}^{\prime\prime\prime)}),
\end{equation}
due to the separation of the probability densities, $p(\kappa(\bmath{\theta}),\kappa(\bmath{\theta}^\prime),\delta) = p(\delta)  p(\kappa(\bmath{\theta}),\kappa(\bmath{\theta}^\prime))$ for the two-point correlation function and $p(\kappa(\bmath{\theta}),\kappa(\bmath{\theta}^\prime),\kappa(\bmath{\theta}^{\prime\prime}),\delta) = p(\delta)  p(\kappa(\bmath{\theta}),\kappa(\bmath{\theta}^\prime),\kappa(\bmath{\theta}^{\prime\prime}))$ for the three-point correlation function and the normalisation condition $\int\dd\delta\:p(\delta)=1$. 
{\robert{Thus one will see the effects of the local density $\delta$ at the observers position if the triangle defined by $p(\kappa(\bmath{\theta}),\kappa(\bmath{\theta}^\prime),\delta)$ is of equilateral shape, or if the distance from the observer to the lenses is small compared to the distance between the lenses on the sky. This is in fact only the case for shallow surveys, alternatively for the redshift bins at low redshift, or large angular scales. This generalises to higher order configurations.}}

\begin{figure}
\begin{center}
\includegraphics[width = .55\textwidth]{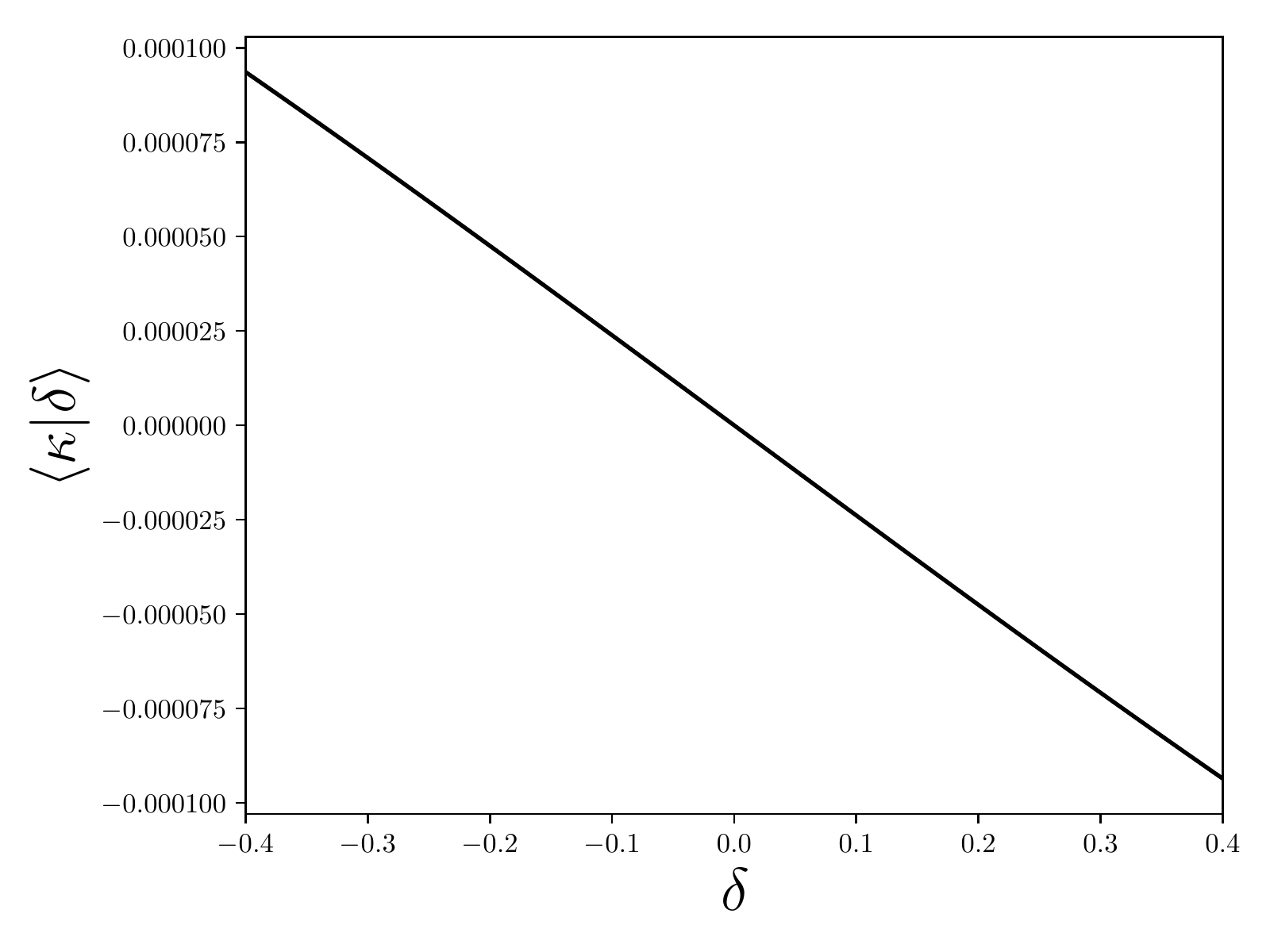}
\caption{Average of the convergence conditionalized on the density $\delta$ at the observer. $\delta=0$ corresponds to standard value, where also the average of $\kappa$ vanishes.}
\label{fig:average_kappa}
\end{center}
\end{figure}

It should be noted that imposing the condition $\delta\neq0$ on the random variable $\delta$ changes already the expectation value of the field $\kappa(\bmath{\theta})$ from zero,
\begin{equation}
\bra\kappa(\bmath{\theta})|\delta\ket =
\int\dd\kappa(\bmath{\theta})\:\kappa(\bmath{\theta})p(\kappa(\bmath{\theta}),\delta),
\end{equation}
if $\delta$ is in fact correlated with $\kappa(\bmath{\theta})$, $\bra\kappa(\bmath{\theta})\delta\ket\neq 0$, even if $\bra\delta\ket$ and $\bra\kappa(\bmath{\theta})\ket$ both vanish. This can be seen in \autoref{fig:average_kappa}, where we show the average convergence as a function of overdensity at the observer's position. {\robert{One clearly sees that observers in underdense regions over estimate the convergence and vice versa for overdense regions.}}

Analysing the structure of the calculation shows a general structure for correlations at arbitrary order: The measurement of a correlation function of order $n$ of a large-scale structure signal requires a joint, $n+1$-variate distribution of the $n$ line of sight-integrated observables and the observer's position. This $n+1$-variate distribution can either be constructed only containing pairwise Gaussian, disconnected terms or using the full non-Gaussian structure, possibly truncated at a specified order.

\subsection{Gaussian and non-Gaussian models for the probability distribution}
The distribution $p(x_i)$ can be assembled from its cumulants, which are analytically computable from a model of structure formation, through the characteristic function $M(t_i)$, which is defined as the Fourier-transform of $p(x_i)$,
\begin{equation}
M(t_j) = \int\prod_i\dd x_i\:p(x_i)\exp\left(-\ci x_i\delta_{ij}t_j\right),\quad\mbox{and}\quad
p(x_i) = \int\prod_j\frac{\dd t_j}{2\pi}\:M(t_j)\exp\left(+\ci x_i\delta_{ij}t_j\right).
\end{equation}
The expansion of $M(t_j)$ in terms of cumulants $K_{i\cdots}$ results in the polynomial
\begin{equation}
\ln\left(M(t_j)\right) = K_it_i + \frac{1}{2!}K_{ij}t_it_j + \frac{1}{3!}K_{ijk}t_it_jt_k +\cdots,
\end{equation}
such that the characteristic function $M(t_j)$ is obtained by exponentiation,
\begin{equation}
M(t_j) = \exp\left(K_it_i + \frac{1}{2!}K_{ij}t_it_j + \frac{1}{3!}K_{ijk}t_it_jt_k +\cdots\right),
\end{equation}
and finally $p(x_i)$ by inverse Fourier transform, where the normalisation condition needs to be imposed in the general case,
\begin{equation}
\int\prod_i\dd x_i\:p(x_i) = 1.
\end{equation}
Here, we have grouped the random variables $\kappa(\bmath\theta)$ for the different lines of sight and $\delta$ into a multivariate vector with components $x_i$. This implies that the first cumulants are given by $K_0=\bra\delta\ket=0$ and $K_1=\bra\kappa(\bmath\theta)\ket=0$. Together with the second cumulants $K_{ij}$, which are assembled from the four possible correlations $\bra\kappa(\bmath{\theta})\kappa(\bmath{\theta}^\prime)\ket$, $\bra\delta^2\ket$ and the cross-cumulants $\bra\kappa(\bmath{\theta})\delta\ket$ and $\bra\kappa(\bmath{\theta}^\prime)\delta\ket$, this defines the Gaussian model, where the cumulant expansion truncates after the quadratic term. Statistical isotropy requires the cross-cumulants $\bra\kappa(\bmath{\theta})\delta\ket$ and $\bra\kappa(\bmath{\theta}^\prime)\delta\ket$ to be equal. 

In addition, the mixed cumulants obey the Cauchy-Schwarz inequality as positivity conditions,
\begin{equation}
\bra\kappa(\bmath\theta)\delta\ket^2 \leq
\bra\delta^2\ket\bra\kappa(\bmath\theta)^2\ket \text{ and }  
\bra\kappa(\bmath{\theta})\kappa(\bmath{\theta}^\prime)\ket^2 \leq \bra\kappa(\bmath{\theta})^2\ket\bra\kappa(\bmath{\theta}^\prime)^2\ket.
\end{equation}
Due to the definition of the fields $\kappa(\bmath\theta)$ and $\delta$, the first order cumulants $K_i$ are zero. A non-Gaussian distribution, which we consider up to third order, would need to be supplied with the set of order three cumulants $K_{ijk}$. It is clear that vanishing cross correlations between convergence $\kappa(\bmath\theta)$ and density $\delta$ will give rise to a separating probability distribution as a consequence of the fact that the cumulant expansion becomes the sum of two independent series.

Specifically, we construct in this way a Gaussian distribution $p(\kappa(\bmath{\theta}),\kappa(\bmath{\theta}^\prime),\delta)$ for the lensing signal in two directions in parallel to the density field, and add to this distribution a non-Gaussian contribution $\bra\kappa(\bmath{\theta})\kappa(\bmath{\theta}^\prime)\delta\ket$ to the covariance. With these distributions it is possible to compute conditional correlation functions $\bra\kappa(\bmath{\theta})\kappa(\bmath{\theta}^\prime)|\delta\ket$ with the assumptions of Gaussian and of non-Gaussian statistics. Furthermore, we set up a non-Gaussian model for the distribution $p(\kappa(\bmath{\theta}),\kappa(\bmath{\theta}^\prime),\kappa(\bmath{\theta}^{\prime\prime}),\delta)$ for conditional three-point correlation functions $\bra\kappa(\bmath{\theta})\kappa(\bmath{\theta}^\prime)\kappa(\bmath{\theta}^{\prime\prime})|\delta\ket$. In order to keep the computations manageable, we will only supply a model of non-Gaussianities of the bispectrum, which we derive in tree-level Eulerian perturbation theory, and will assume that the kurtosis of the fields is dominated by the Gaussian contribution. This is again justified by the restriction to $\ell \leq 100$.

\subsection{Cumulants of the convergence and density fields}

\subsubsection{Gaussian case}
\label{subsec:Gaussian}
The cumulants $K_{ij}$ needed for the Gaussian model result from the variance of the density field $\delta$ with, if the weak lensing convergence $\kappa(\bmath\theta)$ is involved, its line of sight-projection. It should be noted that we will not be in a position to employ the Limber-approximation \citep{loverde_extended_2008, simon_how_2007, limber_analysis_1954, castro_bispectrum_2004, kitching_limits_2017} because modes propagating parallel to the line of sight will be responsible for correlations between the observer's position and the weak lensing signal. Although our primary interest are weak lensing correlation functions and their possible dependence on the density field value of the observer's location, our formalism generalises straightforwardly to correlation functions of other observables or on other properties of the large-scale structure that might characterise an observer.

Firstly, the correlation function $\bra\kappa(\bmath{\theta})\kappa(\bmath{\theta}^\prime)\ket$ of the weak lensing signal can be computed to be
\begin{equation}
\bra\kappa(\bmath{\theta})\kappa(\bmath{\theta}^\prime)\ket =
\int\dd\chi\:W(\chi)\int\dd\chip\:W(\chip) 
\int\frac{\dd^3k}{(2\pi)^3}\:\int\frac{\dd^3k^\prime}{(2\pi)^3}\bra\delta(\bmath{k})\delta(\bmath{k}^\prime)\ket
\exp\left(\ci(\bmath{k}\bmath{x}+\bmath{k}^\prime\bmath{x}^\prime)\right).
\end{equation}
This expression can be simplified by subsituting the spectrum $P(k)$ from eqn. (\ref{eq:powerspectrum})  and performing one trivial integration to obtain
\begin{equation}
\bra\kappa(\bmath{\theta})\kappa(\bmath{\theta}^\prime)\ket = 
\int\dd\chi\:W(\chi)\int\dd\chip\:W(\chip)
 \int\frac{\dd^3k}{(2\pi)^3}\:P(k)\exp\left(\ci\bmath{k}(\bmath{x}-\bmath{x}^\prime)\right)\; ,
\end{equation}
which yields after double substitution of the Rayleigh-expansion for $\exp(\ci\bmath{k}\bmath{x})$ and $\exp(-\ci\bmath{k}\bmath{x}^\prime)$,
\begin{equation}
\exp(\ci\bmath{k}\bmath{x}) =
4\pi\sum_\ell \ci^\ell j_\ell(kx)\sum_m Y_{\ell m}(\hat{k})Y_{\ell m}^*(\hat{x}),
\end{equation}
usage of the orthonormality relation of the spherical harmonics,
\begin{equation}
\int\dd\Omega\:Y_{\ell m}(\hat{k})Y_{\ell^\prime m^\prime}^*(\hat{k}) =
\delta_{\ell\ell^\prime}\delta_{m m^\prime},
\end{equation}
and of the addition theorem of spherical harmonics,
\begin{equation}
\sum_m Y_{\ell m}(\hat{x})Y^*_{\ell m}(\hat{x}^\prime) = \frac{2\ell+1}{4\pi}P_\ell(\cos\gamma),
\end{equation}
the final result
\begin{equation}
\label{eq:kappakappa}
\bra\kappa(\bmath{\theta})\kappa(\bmath{\theta}^\prime)\ket =
\int\dd\chi\:W(\chi)\int\dd\chip\:W(\chip)\sum_\ell(2\ell+1)P_\ell(\cos\gamma)
\int\frac{k^2\dd k}{2\pi^2}\:j_\ell(k\chi)j_\ell(k\chi^\prime)\:P(k).
\end{equation}
Here we use the Legendre-polynomials $P_\ell(\cos\gamma)$ of order $\ell$, the angle $\gamma$ between the directions $\bmath{\theta}$ and $\bmath{\theta}^\prime$ and the spherical Bessel functions $j_\ell(kx)$.

We repeat the calculation for the cross-variance $\bra\kappa(\bmath{\theta}^\prime)\delta\ket$ between the lensing signal $\kappa(\bmath\theta^\prime)$ and the density $\delta$ at the observer's location $\bmath{y}$
\begin{equation}
\bra\kappa(\bmath{\theta}^\prime)\delta\ket = 
\int\dd\chi\:W(\chi)\int\frac{\dd^3k}{(2\pi)^3} 
 \int\frac{\dd^3k^\prime}{(2\pi)^3}\:\bra\delta(\bmath{k})\delta(\bmath{k}^\prime)\ket \exp\left(\ci\bmath{k}\bmath{x}+\bmath{k}^\prime\bmath{y}\right).
\end{equation}
Performing again a trivial integral, by using the orthogonality relation of the exponential function we obtain:
\begin{equation}\label{eq:gauss_kappa_delta}
\bra\kappa(\bmath{\theta}^\prime)\delta\ket =
\int\dd\chi^\prime\:W(\chi^\prime)\sum_\ell(2\ell+1)P_\ell(\cos\gamma)
\int\frac{k^2\dd k}{2\pi^2}\:j_\ell(k\chi)j_\ell(0)\:P(k).
\end{equation}
with the observer positioned at $\chi = 0$ allowed by statistical homogeneity, and where the angular integrations in $k$-space have been carried out using statistical isotropy as before.

Lastly, the variance $\bra\delta^2\ket$ of the density field is straightforwardly computed to be
\begin{equation}\label{eq:gauss_delta_delta}
\bra\delta^2\ket =
\int\frac{\dd^3k}{(2\pi)^3}\:\int\frac{\dd^3k^\prime}{(2\pi)^3}\:\bra\delta(\bmath{k})\delta(\bmath{k}^\prime)\ket\exp\left(\ci(\bmath{k}+\bmath{k}^\prime)\bmath{y}\right) =
\int\frac{k^2\dd k}{2\pi^2}\: P(k),
\end{equation}
all by substitution of the spectrum $\bra\delta(\bmath{k})\delta(\bmath{k}^\prime)\ket = (2\pi)^3\delta_D(\bmath{k}+\bmath{k}^\prime)P(k)$ and subsequent setting of $\bmath{y}$ to zero, as well as carrying out the angular integrations due to statistical isotropy. It is interesting to see how statistical homogeneity and isotropy of the density field reduce the dimensionality of the integrations, in fact $\bra\kappa(\bmath{\theta})\kappa(\bmath{\theta}^\prime)\ket$ can be computed from four instead of eight integrations, $\bra\kappa(\bmath{\theta})\delta\ket$ with two integrations and finally $\bra\delta^2\ket$ from a single integration. In principle, the above listed integrals, for the Gaussian as well as for the non-Gaussian case, would be equally well to deal with inhomogeneous random fields, although at higher computational complexity.

We point out that we can not simplify the integrals further, as it would be commonly done in gravitational lensing: Assuming that there are no correlations in the lensing deflection along the line of sight and assuming that there are no correlations of the lensing deflection occuring at different distances along two lines of sight would set the observer-dependence of the lensing signal to zero.

\subsubsection{Non-Gaussian case}
\label{subsec:nonGaussian}
The non-Gaussian distribution requires at third order the cumulants $K_{ijk}$ in addition to $K_{ij}$. We begin with the three-point correlation function $\bra\kappa(\bmath{\theta})\kappa(\bmath{\theta}^\prime)\kappa(\bmath{\theta}^{\prime\prime})\ket$ of the lensing signal,
\begin{equation}
\bra\kappa(\bmath{\theta})\kappa(\bmath{\theta}^\prime)\kappa(\bmath{\theta}^{\prime\prime})\ket = 
\int\dd\chi\:W(\chi)\int\dd\chip\:W(\chip)\int\dd\chipp\:W(\chipp)
\int\frac{\dd^3k}{(2\pi)^3}\int\frac{\dd^3k^\prime}{(2\pi)^3}\int\frac{\dd^3k^{\prime\prime}}{(2\pi)^3}\bra\delta(\bmath{k})\delta(\bmath{k}^\prime)\delta(\bmath{k}^{\prime\prime})\ket
\exp\left(\ci\bmath{k}\bmath{x}+\bmath{k}^\prime\bmath{x}^\prime+\bmath{k}^{\prime\prime}\bmath{x}^{\prime\prime}\right),
\end{equation}
followed by the three-point correlation function $\bra\kappa(\bmath{\theta})\kappa(\bmath{\theta}^\prime)\delta\ket$ involving $\delta$ at the observer's location,
\begin{equation}
\bra\kappa(\bmath{\theta})\kappa(\bmath{\theta}^\prime)\delta\ket =
\int\dd\chi\:W(\chi)\int\dd\chip\:W(\chip)
\int\frac{\dd^3k}{(2\pi)^3}\int\frac{\dd^3k^\prime}{(2\pi)^3}\int\frac{\dd^3k^{\prime\prime}}{(2\pi)^3}
\bra\delta(\bmath{k})\delta(\bmath{k}^\prime)\delta(\bmath{k}^{\prime\prime})\ket\exp\left(\ci\bmath{k}\bmath{x}+\bmath{k}^\prime\bmath{x}^\prime+\bmath{k}^{\prime\prime}\bmath{y}\right),
\end{equation}
as well as the mixed three-point correlation function $\bra\kappa(\bmath{\theta})\delta^2\ket$,
\begin{equation}
\bra\kappa(\bmath{\theta})\delta^2\ket =
\int\dd\chi\:W(\chi)\int\frac{\dd^3k}{(2\pi)^3}\int\frac{\dd^3k^\prime}{(2\pi)^3}\int\frac{\dd^3k^{\prime\prime}}{(2\pi)^3}
 \bra\delta(\bmath{k})\delta(\bmath{k}^\prime)\delta(\bmath{k}^{\prime\prime})\ket\exp\left(\ci\bmath{k}\bmath{x}+(\bmath{k}^\prime+\bmath{k}^{\prime\prime})\bmath{y}\right),
\end{equation}
and lastly, the skewness $\bra\delta^3\ket$ of the density field at the observer's position which is given by
\begin{equation}
\bra\delta^3\ket = 
\int\frac{\dd^3k}{(2\pi)^3}\int\frac{\dd^3k^\prime}{(2\pi)^3}\int\frac{\dd^3k^{\prime\prime}}{(2\pi)^3}\bra\delta(\bmath{k})\delta(\bmath{k}^\prime)\delta(\bmath{k}^{\prime\prime})\ket
 \exp\left(\ci(\bmath{k}+\bmath{k}^\prime+\bmath{k}^{\prime\prime})\bmath{y}\right).
\end{equation}
Substitution of the bispectrum $\bra\delta(\bmath{k})\delta(\bmath{k}^\prime)\delta(\bmath{k}^{\prime\prime})\ket = (2\pi)^3\delta_D(\bmath{k}+\bmath{k}^\prime+\bmath{k}^{\prime\prime}) B(\bmath{k},\bmath{k}^\prime,\bmath{k}^{\prime\prime})$ for the statistically homogeneous case, using the Dirac distribution for carrying out one of the integrations and setting $\bmath{y}=0$ simplifies the expressions significantly:
\begin{equation}
\bra\kappa(\bmath{\theta})\kappa(\bmath{\theta}^\prime)\kappa(\bmath{\theta}^{\prime\prime})\ket =
\frac{1}{(2\pi)^6}
\int\dd\chi\:W(\chi)\int\dd\chip\:W(\chip)\int\dd\chipp\:W(\chipp)
\int\dd^3k\:\int\dd^3k^\prime\:
B(\bmath{k},\bmath{k}^\prime,-(\bmath{k}+\bmath{k}^\prime))
\exp\left(\ci(\bmath{k}(\bmath{x}-\bmath{x}^\prime) + \bmath{k}^\prime(\bmath{x}^\prime - \bmath{x}^{\prime\prime})\right),
\end{equation}
for the three-point correlation function $\bra\kappa(\bmath{\theta})\kappa(\bmath{\theta}^\prime)\kappa(\bmath{\theta}^{\prime\prime})\ket$ of the lensing signal,
\begin{equation}
\bra\kappa(\bmath{\theta})\kappa(\bmath{\theta}^\prime)\delta\ket =
\frac{1}{(2\pi)^6}
\int\dd\chi\:W(\chi)\int\dd\chip\:W(\chip)
\int\dd^3k\:\int\dd^3k^\prime\:
B(\bmath{k},\bmath{k}^\prime,-(\bmath{k}+\bmath{k}^\prime))
\exp\left(\ci(\bmath{k}\bmath{x}+\bmath{k}^\prime\bmath{x}^\prime)\right),
\end{equation}
and
\begin{equation}
\bra\kappa(\bmath{\theta})\delta^2\ket =
\frac{1}{(2\pi)^6}
\int\dd\chi\:W(\chi)\int\dd^3k\:\int\dd^3k^\prime\:
B(\bmath{k},\bmath{k}^\prime,-(\bmath{k}+\bmath{k}^\prime))
\exp\left(\ci\bmath{k}\bmath{x}\right),
\end{equation}
for the mixed correlation functions $\bra\kappa(\bmath{\theta})\kappa(\bmath{\theta}^\prime)\delta\ket$ and $\bra\kappa(\bmath{\theta})\delta^2\ket$, and
\begin{equation}
\bra\delta^3\ket =
\frac{1}{(2\pi)^6}
\int\dd^3k\:\int\dd^3k^\prime
B(\bmath{k},\bmath{k}^\prime,-(\bmath{k}+\bmath{k}^\prime)) =
\frac{1}{2\pi}\int k^2\dd k\int k^{\prime 2}\dd k^\prime\int\dd\mu^\prime\: B(\bmath{k},\bmath{k}^\prime,-(\bmath{k}+\bmath{k}^\prime)),
\end{equation}
for the skewness $\bra\delta^3\ket$ of the density field. In the expression of all cumulants, application of homogeneity and and isotropy as statistical symmetries greatly reduces the number of integrations, and the above relationships would generalise to higher-order non-Gaussianities in a straightforward way.

Clearly, the magnitude of the observer dependence will first of all be caused by correlations of the type $\bra\kappa(\bmath\theta)\delta\ket$, i.e. by correlations between the lensing large-scale structure where the lensing efficiency function $W(\chi)$ peaks and the density at the observer. This correlation should decrease with increasing source redshifts, and should be small in comparison to that of the weak lensing signal because it depends strongly on correlations at different positions along the line of sight. Supplying $\bra\kappa(\bmath{\theta})\kappa(\bmath{\theta}^\prime)\delta\ket$ as a non-Gaussian contribution should not change the result dramatically because the covariance of the distribution $p(\kappa(\bmath{\theta}),\kappa(\bmath{\theta}^\prime),\delta)$ is dominated by the Gaussian variances: In this way, the conditional two-point correlation function $\bra\kappa(\bmath\theta)\kappa(\bmath\theta^\prime)|\delta\ket$ would primarily result from correlations between three points and these would be non-existent in linear structure formation and would only be small on large scales.


We would argue that the situation is different for conditional three-point correlation functions $\bra\kappa(\bmath{\theta})\kappa(\bmath{\theta}^\prime)\kappa(\bmath{\theta}^{\prime\prime})|\delta\ket$. It originates from the distribution $p(\kappa(\bmath{\theta}),\kappa(\bmath{\theta}^\prime),\kappa(\bmath{\theta}^{\prime\prime}),\delta)$ whose covariance contains non-diagonal terms reflecting correlations between four points, which are nonzero even for linear and Gaussian statistics. In addition, the distribution of $\delta$ at the location of the observer is empirically described by a lognormal distribution in the non-Gaussian regime of structure formation, such that the average $\bra\delta\ket=0$ but the most likely $\delta$ would become negative, in a volume-averaged sense. And one would need to take biasing into account, because the formation galaxies from where observations would of course need to take place, are associated with high-density regions, in a biasing model beyond linear bias.

\begin{figure}
\begin{center}
\includegraphics[width=8.8cm,height=8.cm,trim={0.3cm 0.4cm 0.4cm 0.3cm},clip]{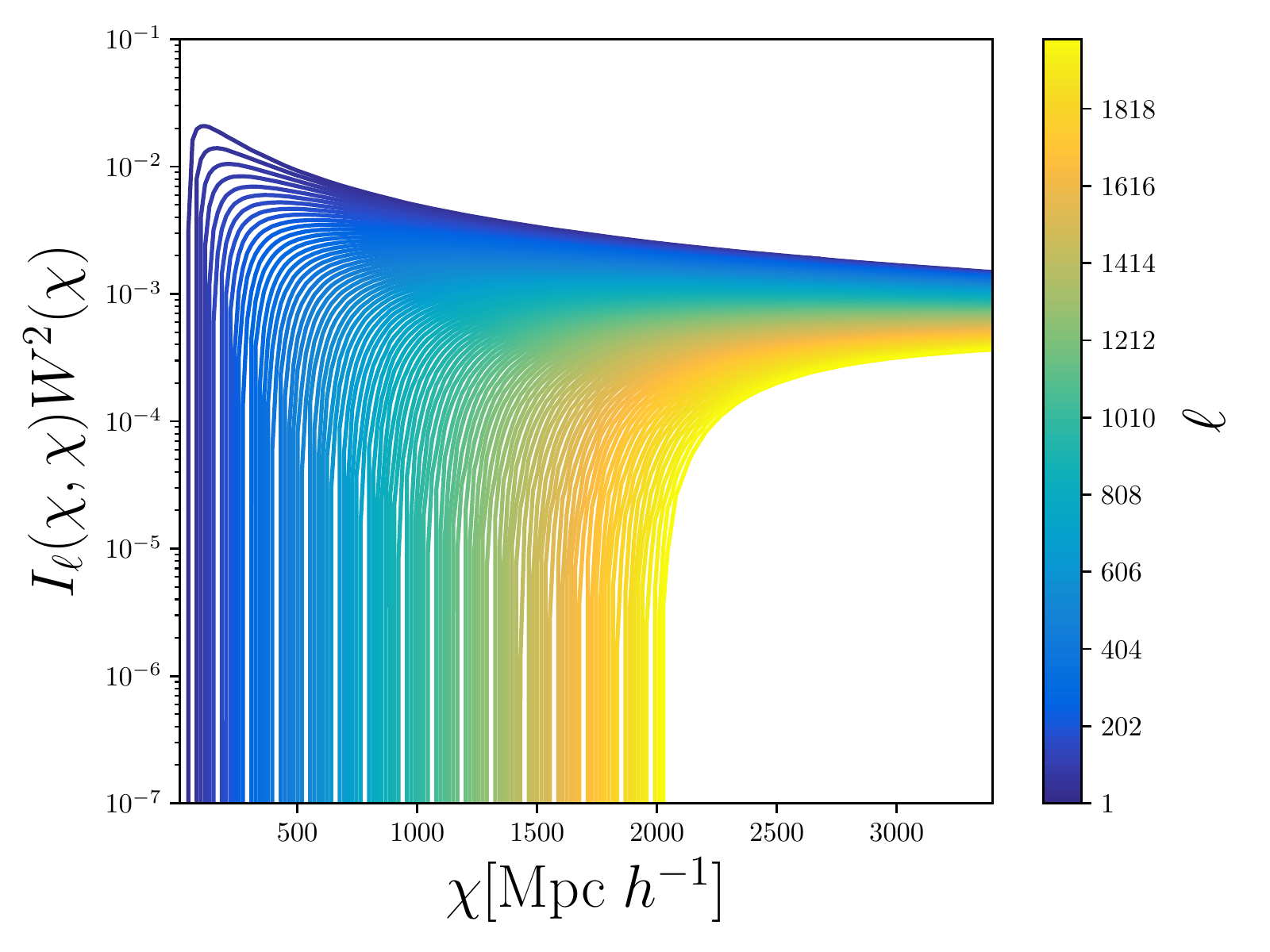}
\includegraphics[width=8.8cm,height=8.cm,trim={0.3cm 0.4cm 0.4cm 0.3cm}]{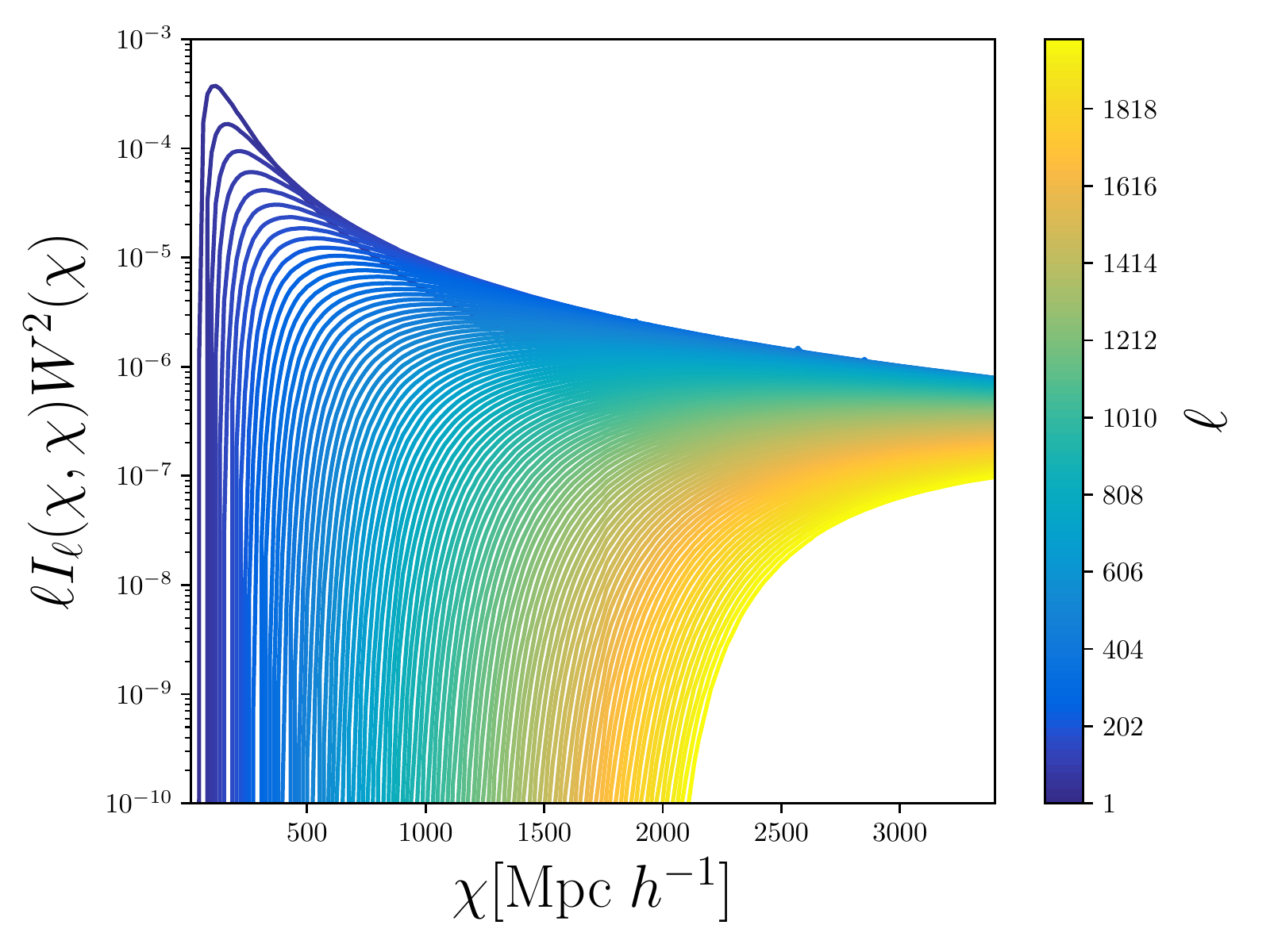}
\caption{{\robert{On the left side we show the diagonal elements, $\chi = \chi^\prime$ of the $k$ integral, $I_\ell(\chi,\chi^\prime)$, in Eq. (\ref{eq:kappakappa}) for different $\ell$-modes as a color bar multiplied with the square of the weight function. The right panel shows it multiplied with the multipole $\ell$ to illustrate the actual contribution of the different modes.}}}
\label{fig:integral}
\end{center}
\end{figure}

\section{Implementation and numerical results}\label{sect_implementation}
In this paper we explicitly implement the Gaussian case described in Sect. \ref{subsec:Gaussian}. For the non-Gaussian case one would need to evaluate highly oscillatory over many dimensions (cf. Sect. \ref{subsec:nonGaussian}). These integrals would be used as the coefficients for the cumulant-expansion which in turn needs to be transformed via fast Fourier-transform methods to get the actual distribution function $p(\kappa(\bmath{\theta})\kappa(\bmath{\theta}^\prime),\delta)$. The situation {\robert{exacerbates}} with increasing order of the correlation functions: Since these integrals are not stable enough numerically we cannot expect to obtain a reliable estimate of the effect we are seeking for in the non-Gaussian case, but we will provide estimates and a discussion of the influence of non-Gaussian effects.

Throughout this section we will assume a survey with \textit{Euclid}'s redshift distribution (\ref{eq:srd}) and a single tomographic redshift bin{\robert{, i.e. we integrate over the whole source distribution}}. The technical challenges are integrals over Bessel-weighted functions, which is required by the fact that correlations between the fluctuations in the density field that are mapped out by gravitational lensing and the density field value at the observer's location need to be represented and imposing the Limber-approximation, which neglects line of sight-correlations of the signal, is not permissible. From a Gaussian model of the conditionalised correlation function we are able to derive the effect of local structures onto the weak lensing signal.

\subsection{Conditionalised correlation function}
For the conditionalised two point correlation functions we assume the distribution $p(\kappa(\bmath{\theta})\kappa(\bmath{\theta}^\prime),\delta)$ in eqn. (\ref{eq:spectra}) to be a Gaussian distribution. On sufficiently large scales we expect the influence of non-Gaussianities to be small compared to the Gaussian part. We thus have
\begin{equation}
p(\kappa(\bmath{\theta})\kappa(\bmath{\theta}^\prime),\delta) = \frac{1}{\sqrt{(2\pi)^3\mathrm{det}\bmath{C}}}\exp \left( -\frac{1}{2}\bmath{X}^T\bmath{C}^{-1} \bmath{X}\right)\; ,
\end{equation}
where
\begin{equation}
\bmath{X}^T = (\kappa(\bmath{\theta})\kappa(\bmath{\theta}^\prime),\delta)
\end{equation}
and
\begin{equation}
\label{eq:covariance}
\bmath{C} = 
\begin{pmatrix}
\langle\kappa\kappa \rangle & \langle\kappa\kappa^\prime\rangle & \langle\kappa\delta\rangle \\ 
\langle\kappa\kappa^\prime\rangle & \langle\kappa^\prime\kappa^\prime \rangle &
\langle\kappa^\prime\delta \rangle \\
\langle\kappa\delta\rangle & \langle\kappa^\prime\delta \rangle & \langle\delta\delta\rangle
\end{pmatrix}.
\end{equation}

As discussed before, the observer dependence is sourced by the correlations $\langle\kappa\delta\rangle$ and $\langle\kappa^\prime\delta\rangle$. Furthermore, the dependence on the angle $\gamma\equiv |\bmath{\theta} - \bmath{\theta}^\prime|$ is dictated by the correlator $\langle\kappa^\prime\delta\rangle$ only.

As pointed out, the main challenge is to integrate the spherical Bessel-functions as in eqn.~(\ref{eq:kappakappa}). We do this using the Levin-integration \cite{levin_fast_1996,levin_analysis_1997} already implemented and tested in \cite{zieser_cross-correlation_2016,spurio_mancini_3d_2018}. Once set up for the specific problem it can calculate integrals over rapidly oscillatory functions. Relevant for us are integrals of the type:
\begin{equation}
I_\ell (\chi,\chi^\prime) \equiv \int\frac{k^2\mathrm{d}k}{2\pi^2}j_\ell(k\chi) j_\ell(k\chi^\prime) P(k)\;,
\end{equation}
which occur in the projection of spectra without the assumption of Limber's approximation. We see that the diagonal dominates the integral, especially for high $\ell$, this feature shows exactly why Limber's approximation is so powerful \citep{kitching_limits_2017}. \autoref{fig:integral} shows the diagonal entries of the same integral on the left from $\ell = 1$ up to $\ell = 2000$. It can be clearly seen how the peak moves to higher $\chi$ and turns into a plateau as $\ell$ increases. Furthermore, the integral stays stable for very high $\ell$ and $\chi$.

More importantly, the right plot shows the integral multiplied with the weight function of the lens mapping in eqn. (\ref{eq:lens_mapping}). This is effectively the quantity entering in eqn. (\ref{eq:kappakappa}) and it shows the importance of the lower $\ell$-modes compared to the higher ones when calculating the sum in eqn.~(\ref{eq:kappakappa}). For the rest of this work we will stop the sum at $\ell = 2000$. Note that this remains true even with the additional factor $(2\ell +1)$ since the weight function suppress the contribution from large $\chi$.

\begin{figure}
\begin{center}
\includegraphics[width=8.8cm,height=8.cm,trim={0.3cm 0.4cm 0.4cm 0.3cm},clip]{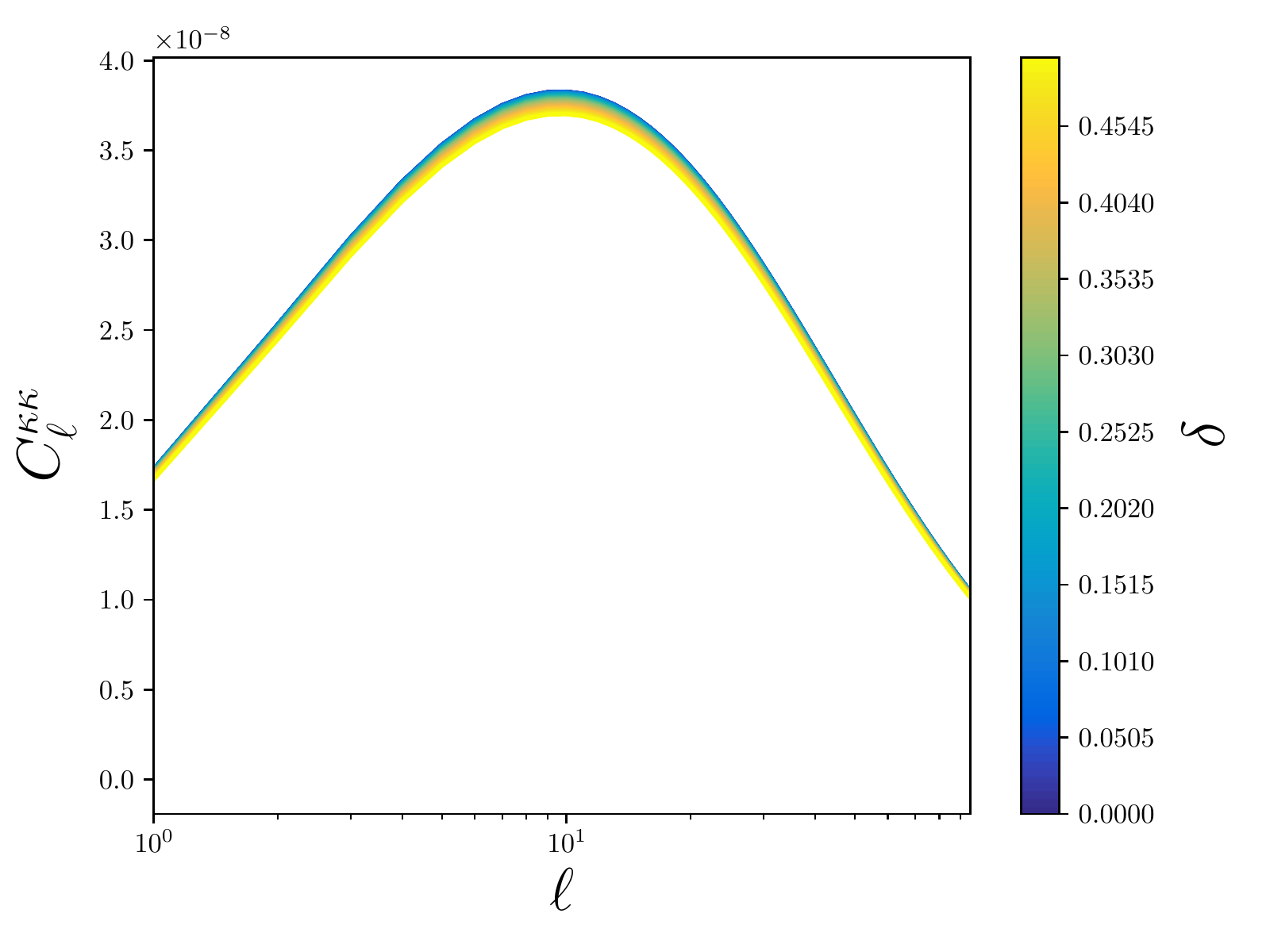}
\includegraphics[width=8.8cm,height=8.cm,trim={0.3cm 0.4cm 0.4cm 0.3cm},clip]{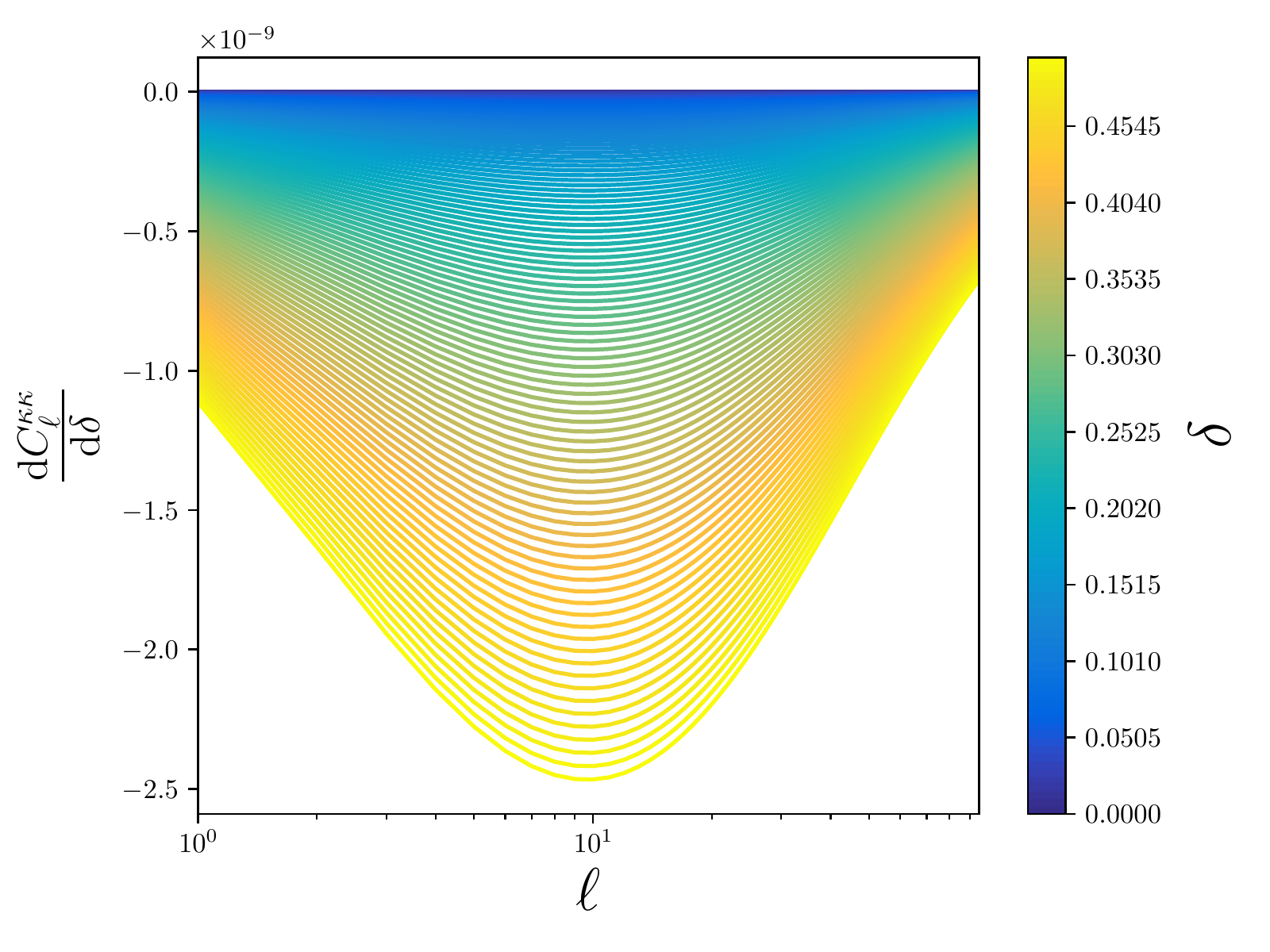}
\caption{\textit{Left}: Convergence angular power spectra for a single tomographic bin in the multipole range $\ell = 1$ to $\ell = 100$ for observers situated at overdense regions. \textit{Right}: Change of the convergence angular power spectrum with respect to the overdensity at the observer. In both plots the overdensity is colour-coded. {\robert{It should be noted that the same results would be obtained for negative $\delta$ since the situation is completely symmetric in the Gaussian case, see \autoref{fig:spectra_delta_single}.}}} 
\label{fig:spectra_delta}
\end{center}
\end{figure}

\begin{figure}
\begin{center}
\includegraphics[width=8.8cm,height=8.cm,trim={0.3cm 0.4cm 0.4cm 0.3cm},clip]{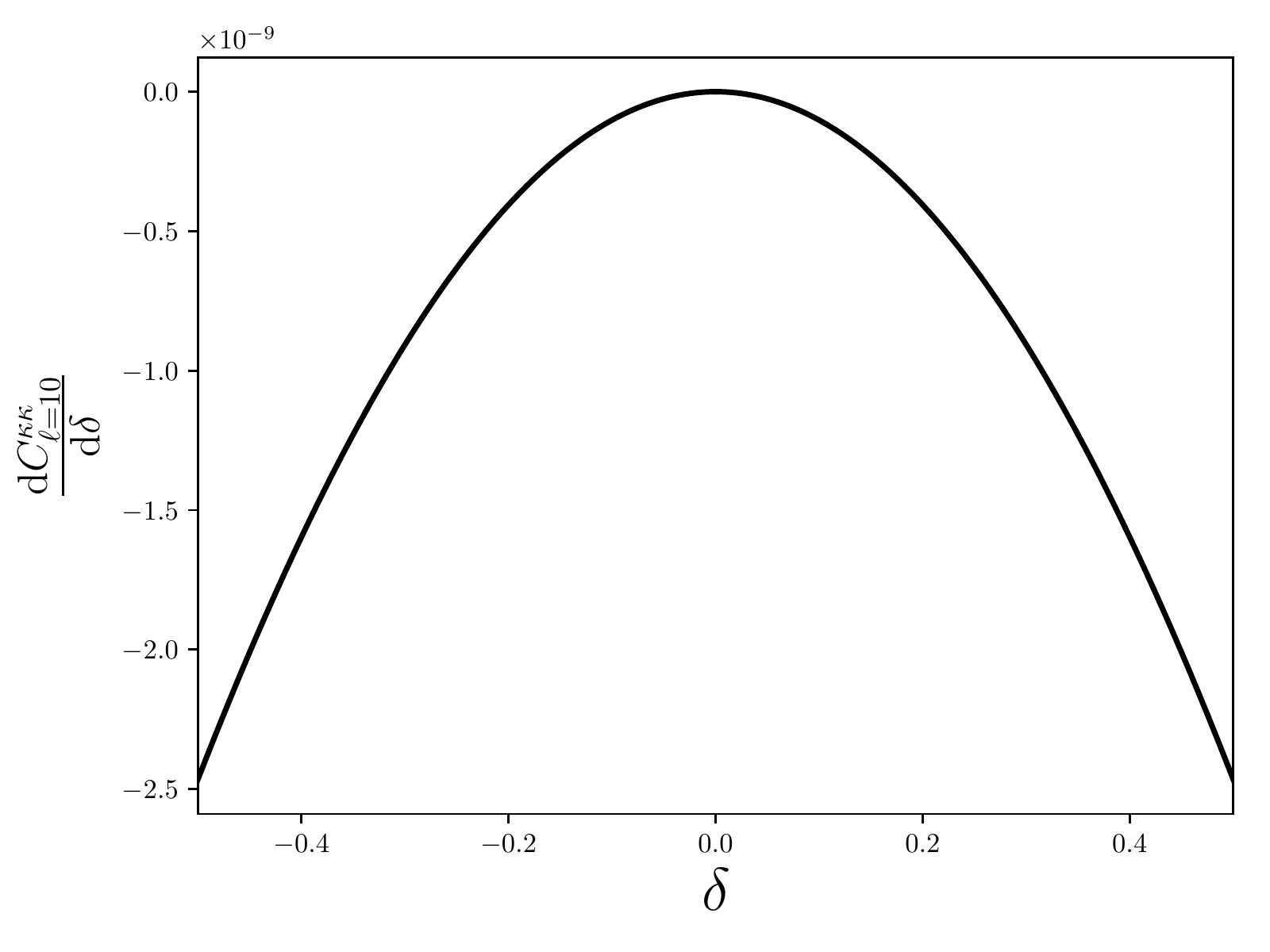}
\caption{{\robert{Change of the convergence angular power spectrum with respect to the overdensity at the observer as a function of the overdensity at a single multipole $\ell = 10$.}}}
\label{fig:spectra_delta_single}
\end{center}
\end{figure}

\subsection{The effect of the local Universe}
Before we continue, a short discussion about the expected differences between the Gaussian and non-Gaussian case is necessary.
In the Gaussian case all cumulants entering in eqn. (\ref{eq:spectra}), i.e. only two points will be correlated. The observer-dependence is driven by the cross-correlations $\langle\kappa^\prime\delta\rangle$ and $\langle\kappa\delta\rangle$. However, both expressions (\ref{eq:gauss_kappa_delta}) and (\ref{eq:gauss_delta_delta}) are similar to eqn. (\ref{eq:kappakappa}), except for the line of sight integration and the argument in the bessel function, we do not expect any dependence on the angle $\gamma$. That is, the correlation function (\ref{eq:spectra}) will of course depend on the angle $\gamma$, however, the correction due to a non-vanishing $\delta$ will be the same on all scales. It therefore suffices to evaluate eqn. (\ref{eq:spectra}) at an arbitrary angle and rescale the correlation function accordingly.
This is expected to change when including non-Gaussian terms in the probability distribution, because the mode coupling function (\ref{eq:mode_coupling_f2}) will prefer certain configurations of wave vectors, which will couple the radial distance to the angle. However, these contributions are small compared in the Gaussian case as already pointed out, since on scales where the expansion in cumulants is applicable all variances are dominated by their Gaussian contribution. For higher order statistics such as the bispectrum the situation can become differently on smaller scales, since the intrinsic contribution of the Gaussian part vanishes. 

In the left panel of \autoref{fig:spectra_delta} we show the weak lensing spectrum $C^{\kappa\kappa}_\ell$ derived from the correlation function (\ref{eq:spectra}) for different values of $\delta$ at the observer's position. The latter is represented by the fraction of the cosmic volume with this density or larger for a Gaussian random field. We restrict the multipole range to $\ell \leq 100$ for linear theory to be applicable. The plot is restricted to overdense regions, it should however be noted that the situation is completely symmetric in the Gaussian case, i.e. the same would hold for an observer in an underdense region with higher observed amplitudes for the spectrum. {\robert{This can be clearly seen in \autoref{fig:spectra_delta_single}, where we show the change of the power spectrum with the overdensity as a function of the overdensity itself. }}
One can clearly see that observations from overdense regions being exposed to less convergence induced by the LSS. Intuitively, this can be understood as a consequence of the normalisation condition of the correlation function: $\int\dd^3x\:C(r)=0$, which needs to be fulfilled along each line of sight. If the observer is placed in a region with high amplitudes and consequently large variance, the variance of the field being observed is going to be smaller.

The right panel of the same figure shows how the convergence power spectrum changes as a function of the multipole order and via a colour bar also again as a function of the density at the observer's position. It is important to note that the point $\delta = 0$, or unity in the colour bar is a maximum, this is clear from the fact that the situation is completely symmetric around $\delta = 0$, where the probability function peaks. Furthermore, we note that the derivative increases the farther we go away from $\delta =0$. The general shape of the curve resembles the shape of the power spectrum in the right panel.

\begin{figure}
\begin{center}
\includegraphics[width=8.8cm,height=8.cm,trim={0.3cm 0.4cm 0.4cm 0.3cm},clip]{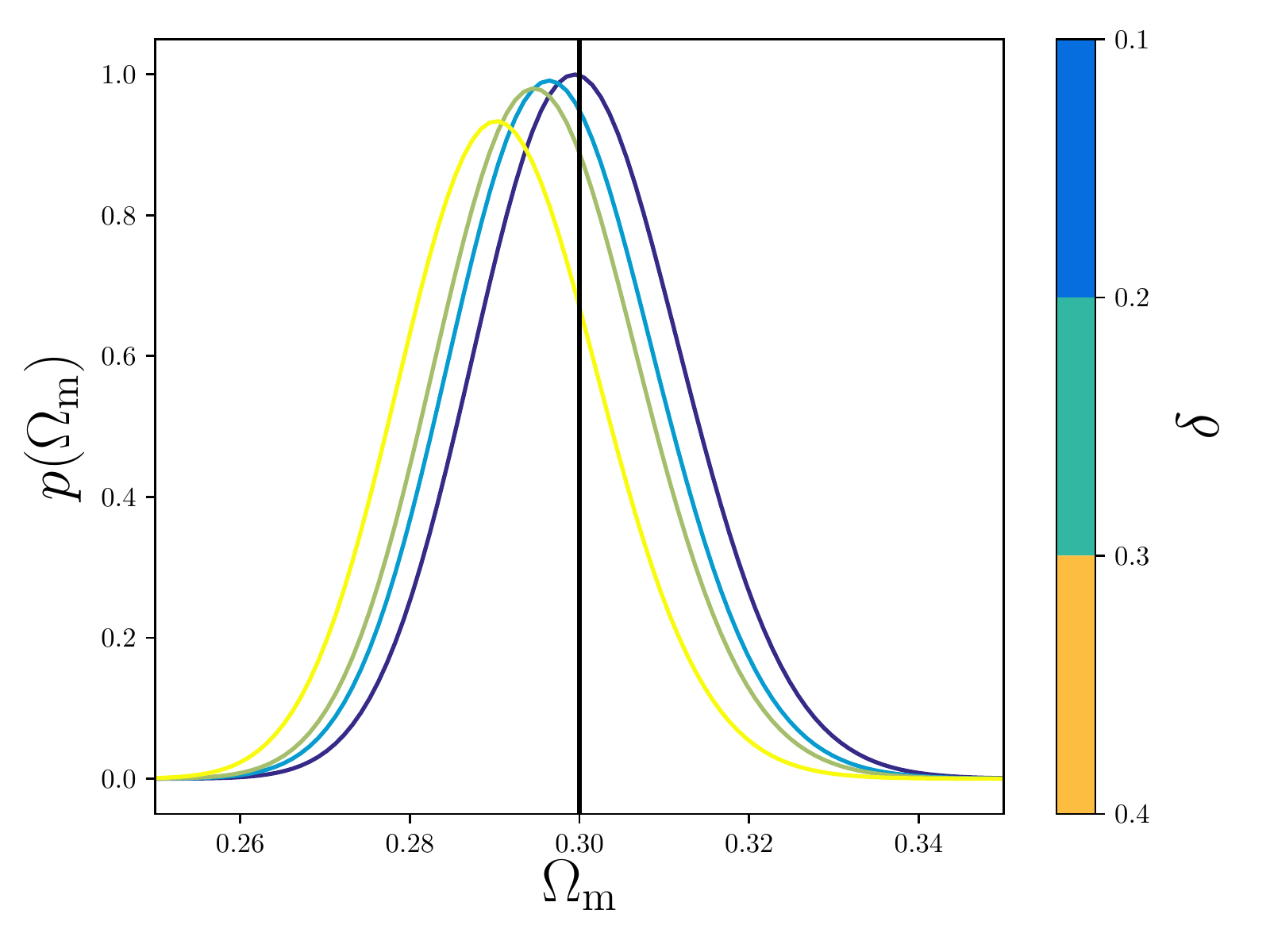}
\includegraphics[width=8.8cm,height=8.cm,trim={0.3cm 0.4cm 0.4cm 0.3cm},clip]{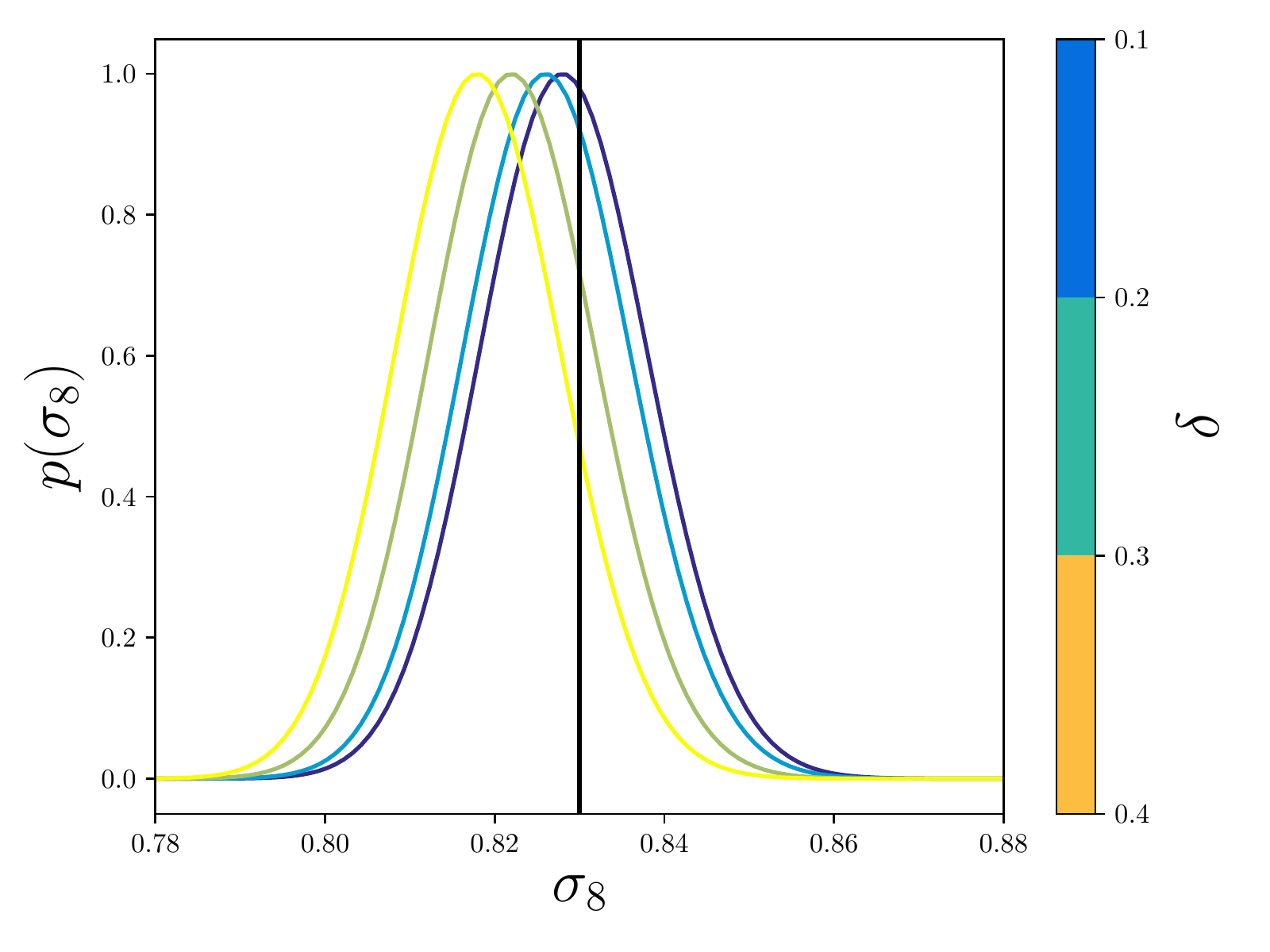}
\caption{Posterior distribution for $\Omega_\mathrm{m}$ (\textit{left}) and $\sigma_8$ (\textit{right}) as a function of the overdensity at the observer's position. Only multipoles up to $\ell=100$ are considered, in order to be sensitive to linear, Gaussian structures only.}
\label{fig:likelihood} 
\end{center}
\end{figure}

As a simple illustration of the effects investigated so far we collect all multipoles below $\ell = 100$, we then fit a single cosmological parameter with a model which does not contain the observer dependent effects:
\begin{equation}
\chi^2 = f_\mathrm{sky}\sum_\ell\frac{2\ell+1}{2} \left[\log\frac{\det C_\ell}{\det\hat{C}_\ell}+ \mathrm{tr}\;\hat{C}_\ell C^{-1}_\ell-1\right]\; .
\end{equation}
Such that $C_\ell$ is the usual convergence power spectrum and $\hat{C}_\ell$ is the power spectrum including observer dependent contributions at the fiducial cosmology. The resulting plots are shown in \autoref{fig:likelihood} for $\Omega_\mathrm{m}$ on the left and $\sigma_8$ on the right.

We see that a biased is induced on the two parameters with an {\robert{observer}} at an overdense region to lower values of both $\sigma_8$ and $\Omega_\mathrm{m}$. The bias is clearly bigger for $\sigma_8$ since it solely determines the amplitude of the lensing spectrum without changing the expansion history of the Universe and thus the conversion from redshift space to comoving distance. It can therefore mimic exactly the influence of an overdense region at the observer. However, in both cases the induced bias lies well in the 1$\sigma$ region of the posterior. Furthermore, collecting more multipoles we would expect the effect of the local density to shrink. The reason being that for the power spectra conditionalized on $\delta$ we need to evaluate a triangle configuration as described by the covariance in eqn. (\ref{eq:covariance}). Moving to smaller angular scales the correlation $\langle\kappa\kappa^\prime\rangle$ will increasingly dominate the correlation $\langle\delta\kappa^\prime\rangle$ in a non-Gaussian model, which would be necessary in this case. Thus the observed spectra will be uncorrelated with the observer's densities at high multipoles, allowing for a separation of the joint probability $p(\kappa,\kappa^\prime,\delta) = p(\kappa,\kappa^\prime)p(\delta)$ resulting into no effect on the spectra. Another influence will come when considering tomographic data. Here the influence will be largest for tomographic bins at low redshifts, since the lensing efficiency function will peak at lower comoving distances, thus increasing the correlation $\langle\kappa^\prime\delta\rangle$. 
{\robert{However, the results here can be seen as, to first order, an effect averaged over all tomographic bins.}}
The magnitude of the biasing effect would decrease if more multipoles beyond $\ell=100$ were included, and from this point of view we can deny that the much-discussed low values for $\Omega_m$ typical for lensing in comparison to the CMB are caused by this effect, besides the fact that $\sigma_8$ is biased as well, in contrast to this particular tension between lensing and CMB-data, so that we consider these two issues as unrelated.

{\robert{The results presented here can also be applied to other survey settings. As described before we expect larger effects for shallower surveys. Thus for a survey like the Dark Energy Survey \citep[DES,][]{DES2005} we expect a larger influence of the local Universe. One finds that in this case the effect on the spectra is roughly 1.5 as big as the one presented here.
However, one should keep in mind that even though the effect at a single multipole, DES has much less access to small multipoles, where the effect is more important,  compared to Euclid. Thus the total effect on the inference process is a trade-off between these to components.}}

In general, any statistical observer-dependence of a cosmological observable would decrease with increasing survey volume. For our estimates we have restricted ourselves to the Euclid survey and only considered low multipoles, but this restriction does not automatically imply Gaussianity. In fact, the fully evolved large-scale structure shows strong non-Gaussianities even at low multipoles, and we would argue that statistical correlations between the observables and the observer location will then become much more involved and possibly stronger.

\begin{figure}
\begin{center}
\includegraphics[width=8.8cm,height=8.cm,trim={0.3cm 0.4cm 0.4cm 0.3cm},clip]{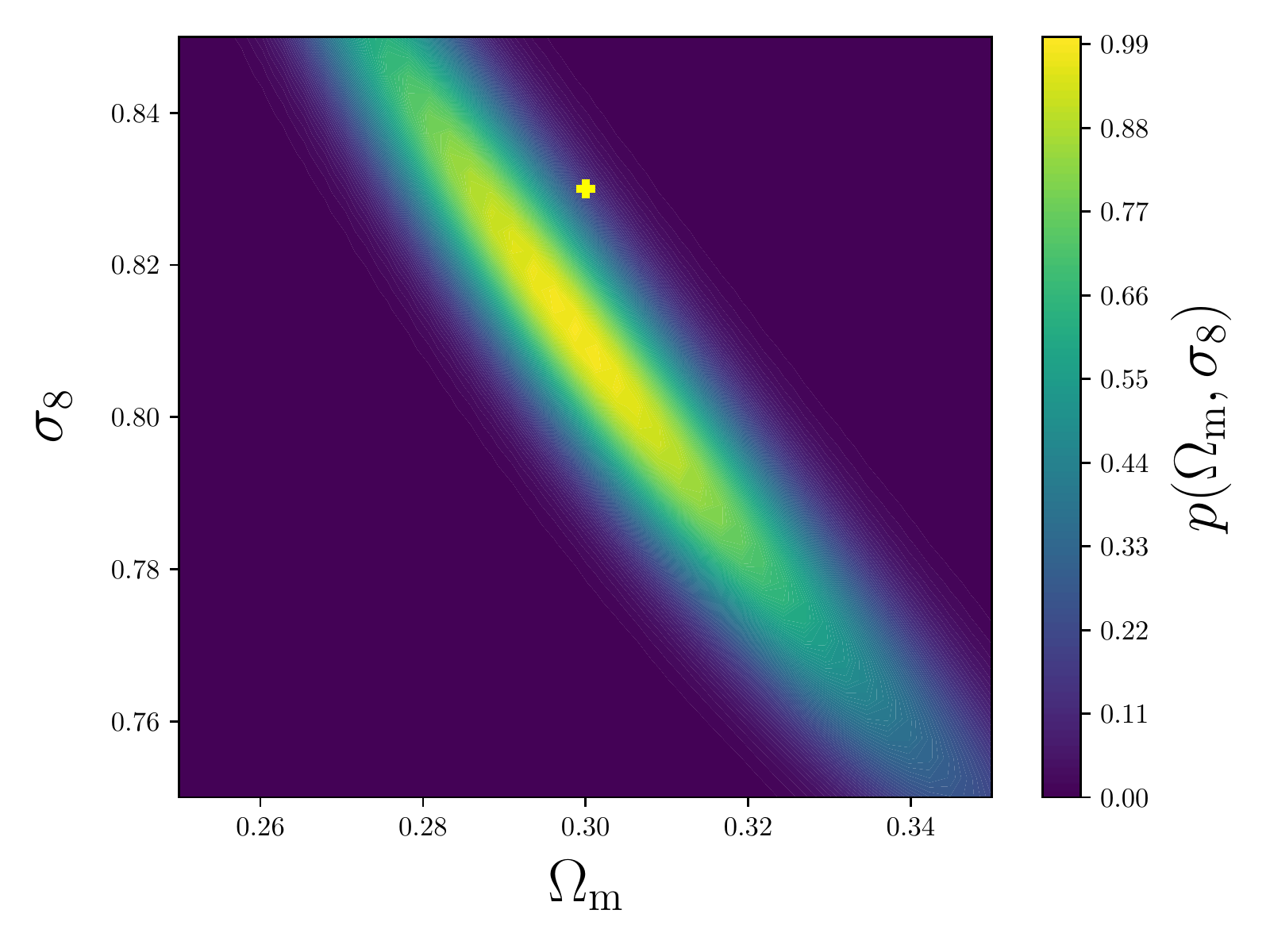}
\caption{Joint posterior distribution in the $(\Omega_\mathrm{m},\sigma_8)$-plane. The survey settings are the same as for the one in \autoref{fig:likelihood}. Here the overdensity at the observer was chosen to be $\delta = 0.4$. The fiducial cosmology is marked with a yellow cross.}
\label{fig:2dc}
\end{center}
\end{figure}

\subsection{Direct simulation of the local Universe's influence}
Depending on the density field at the observer's location, typical changes to the weak lensing spectra are of the order $10^{-2}$, which is difficult to resolve in a numerical simulation. The number $n$ of realisations that would be needed to see a difference $\Delta C_\ell = \hat{C}_\ell-C_\ell$ of the spectrum $C_\ell$ in the presence of cosmic variance given by $\mathrm{cov} = 2/(2\ell+1)C_\ell^2$ for a full-sky observation can be estimated with the Chebyshev-inequality: The probability $p$ that the difference between an estimate of the spectrum $C^\mathrm{est}_\ell$ and the true value $C_\ell$ exceeds a value of $\Delta C_\ell$ is bounded by the covariance,
\begin{equation}
p\left(\left|C^\mathrm{est}_\ell-C_\ell\right| \geq \left|\hat{C}_\ell-C_\ell\right|\right) \leq 
\frac{1}{n}\frac{\mathrm{cov}}{(\hat{C}_\ell-C_\ell)^2}.
\end{equation}
Setting the probability to $p = \mathrm{erfc}(1/\sqrt{2})\simeq0.32$ for a Gaussian $1\sigma$ confidence interval yields an estimate for the number of realisations,
\begin{equation}
n \simeq \frac{1}{\mathrm{erfc}(1/\sqrt{2})} \frac{2}{2\ell+1} \left(\frac{C_\ell}{\hat{C}_\ell-C_\ell}\right)^2,
\end{equation}
where we have substitute the standard expression for the covariance for estimates of an angular spectrum.

With typical changes $\hat{C}_\ell-C_\ell\simeq10^{-2}$ one would then need $n=3000, 1000, 300$ realisations at $\ell=10, 30, 100$, which are prohibitively large numbers. The number of realisations $n$ needed would scale inversely with the sky fraction $f_\mathrm{sky}$. In a related project we have computed realisations of the weak lensing sky by tracing photons through density fields \citep{carbone_lensed_2009, jain_ray-tracing_2000, li_semi-analytic_2011, killedar_gravitational_2012, becker_calclens_2013, antolini_n-body_2014} while being evolved by means of second-order Lagrangian perturbation theory, and we show one of these simulations in Figure~\ref{fig:realisation}, where a clear signature of the local Universe can be seen. The simulation was run in a cube with a side length of $1024~\mathrm{Mpc}/h$, and photon trajectories converging on the observer at the centre of the cube were computed starting from a spherical surface with the radius of $512~\mathrm{Mpc}/h$ while the simulation was running in order to conserve the light cone condition $\dd s^2 = c^2\dd\eta^2-\dd\chi^2=0$ between comoving distance $\chi$ and conformal time $\eta$, where the Born-approximation discards distortions of the past light cone \citep{borzyszkowski_liger_2017}.

\section{Summary}\label{sect_summary}
The subject of this paper is the observer-dependence of cosmological observations, in particular the weak lensing signal. Clearly, there will be deviations between a realistic observer and an idealised FLRW-observer in their state of motion, and the structures that a realistic observer can see are necessarily correlated with the structure in which she or he is residing. We test how the position of the observer within the cosmic environment affects the results of the analysis of the weak lensing signal. Our main result is that there is a slight dependence of the weak lensing spectra and the resulting constraints on cosmological parameters on the observer's position parameterised with the density contrast $\delta$.

The central point of our analytical argumentation is a multivariate distribution which incorporates the joint distribution of the weak lensing signal $\kappa(\bmath\theta)$ in different directions $\bmath\theta$ and the local density $\delta$. To this purpose, we construct a Gaussian distribution incorporating correlations $\bra\kappa(\bmath\theta)\delta\ket$ in its covariance matrix and extend this model for the distribution function to include non-Gaussian structures at the level of three-point correlation functions, which we source from non-linear structure formation with Euler-perturbation theory at lowest order. By constructing distribution $p(\kappa(\bmath{\theta}),\kappa(\bmath{\theta}^\prime),\delta)$ we calculate two-point correlation functions through the second moments with a condition imposed on $\delta$.

We find that the typical change of the weak lensing  power spectra due to observer's $\delta$ is of order of
a few percent, as presented in Fig. \ref{fig:spectra_delta}. Such a change in a power spectra results in a percent-level bias in the inferred cosmological parameters. The bias in the cosmological parameters $\Omega_m$ and $\sigma_8$ results in these parameters being underestimated, cf. Fig.~\ref{fig:likelihood}. This is a very interesting result {\robert{which}} could explain the discrepancy between the measurements of the parameters $\Omega_m$ and $\sigma_8$
inferred from weak {\robert{lensing}} and the CMB constraints, cf. Fig.~\ref{fig:2dc} of this paper and Fig. 10 of \citet{PhysRevD.98.043526}. More studies are required in order to confirm whether the observer-dependence could be the solution to the tension between the weak lensing and CMB constraints, but the difference in direction in parameter space makes this explanation unlikely.

\begin{figure}
\begin{center}
\includegraphics[width = .95\textwidth, trim={0 0 0 0.5cm},clip]{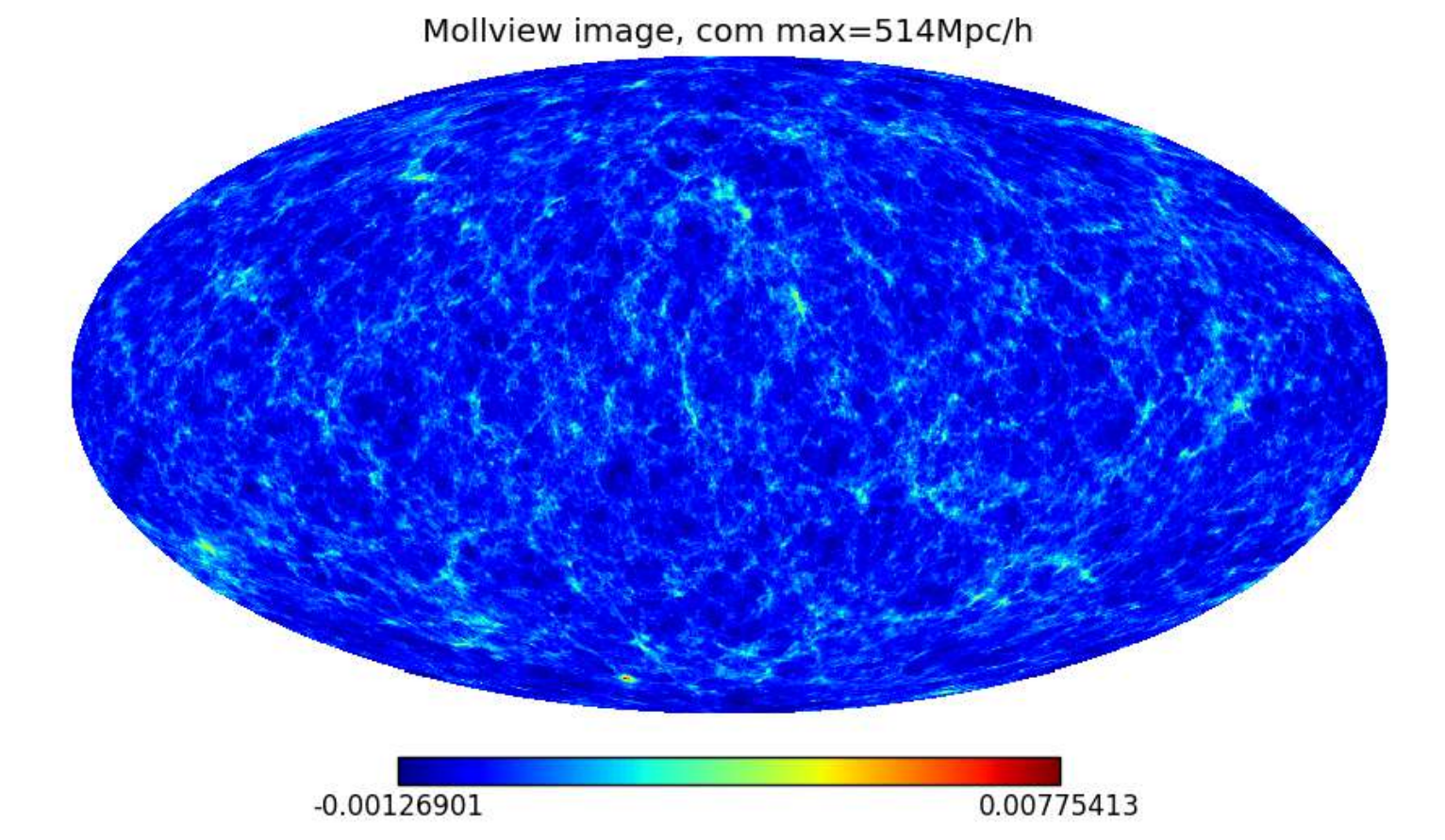}
\caption{All-sky map of the weak lensing convergence from second-order Lagrangian perturbation theory for a $\Lambda$CDM-cosmology. The side-length of the simulation cube is $1024~\mathrm{Mpc}/h$ and the distance to the lensed galaxies is $512~\mathrm{Mpc}/h$.}
\label{fig:realisation} 
\end{center}
\end{figure}

Future extension of this work will include numerical simulations of large-scale structure formation These will be the  constrained simulations with the subsequent reconstruction of weak lensing light cones, conditional on the location of the observer, cf. Fig. \ref{fig:realisation}. This will allow us to investigate the observer-dependence of other cosmological probes and the sensitivities of alternative statistical measures. These results will be important when analysing the future surveys such as the Euclid, where the percent-level precision is required.

\section*{Acknowledgements}
We acknowledge funding by the Australia-Germany Joint Research Cooperation Scheme 2017/2018 by Universities Australia and the German Academic Exchange Service. BMS would like to thank the University of Sydney and the University of Auckland for their hospitality. KB acknowledges the support of the Australian Research Council through the Future Fellowship FT140101270.

\bibliographystyle{mnras_openaccess}
\bibliography{references}

\bsp
\label{lastpage}
\end{document}